\documentclass[preprint]{aastex}
\usepackage{emulateapj5}

\newcommand{\Msol}{M$_{\odot}$}
\newcommand{\Mjup}{M$_{\mathrm{JUP}}$}
\newcommand{\Rjup}{R$_{\mathrm{JUP}}$}
\newcommand{\Msini}{M\,$\sin i$}

\newcommand{\kms}{km\,s$^{-1}$}
\newcommand{\ms}{m\,s$^{-1}$}

\received{13 October 2000}
\accepted{1 December 2000}
\journalid{000}{April 2001}
\slugcomment{To appearin  {\em The Astrophysical Journal, Letters}.}

\lefthead{Tinney et al.}
\righthead{First Results from the Anglo-Australian Planet Search}

\begin{document}
\title{First Results from the Anglo-Australian Planet Search -- A Brown Dwarf Candidate and a 51\,Peg-like Planet\altaffilmark{1} }

\author{C.G. Tinney\altaffilmark{2}, R. Paul Butler\altaffilmark{3,2}, 
        Geoffrey W. Marcy\altaffilmark{4,5}, Hugh R.A. Jones\altaffilmark{6}, 
        Alan J. Penny\altaffilmark{7}, Steven S. Vogt\altaffilmark{8}, 
        Kevin Apps\altaffilmark{9}, Gregory.W. Henry\altaffilmark{10}}

\altaffiltext{1}{Based on observations obtained at the
    Anglo--Australian Telescope, Siding Spring, Australia; 
    and the W.M. Keck Observatory, which is operated jointly
    by the University of California and the California Institute
    of Technology.}
\altaffiltext{2}{Anglo-Australian Observatory, PO Box 296, Epping. 1710. 
Australia. {\tt cgt@aaoepp.aao.gov.au}}
\altaffiltext{3}{Carnegie Institution of Washington,Department of Terrestrial Magnetism,
       5241 Branch Rd NW, Washington, DC 20015-1305}
\altaffiltext{4}{Department of Astronomy, University of California, Berkeley, CA, 94720}
\altaffiltext{5}{Department of Physics and Astronomy, San Francisco State University, San Francisco, CA 94132.}

\altaffiltext{6}{Astrophysics Research Institute, Liverpool John Moores University, 
       Twelve Quays House, Egerton Wharf, Birkenhead CH41 1LD, UK}

\altaffiltext{7}{Rutherford Appleton Laboratory, Chilton, Didcot, Oxon OX11 0QX, U.K.}

\altaffiltext{8}{UCO/Lick Observatory, Unversity of California at Santa Cruz, CA 95064.}

\altaffiltext{9}{Physics and Astronomy, University of Sussex, Falmer, BN1 9QJ, U.K.}

\altaffiltext{10}{Center of Excellence in Information Systems, Tennessee State University, Nashville, TN 37203-3401.}

\begin{abstract}
We report results from the Anglo-Australian Planet Search -- a survey for
planets around 200 solar-type stars in the southern hemisphere, 
which is being carried out on the 3.9m Anglo-Australian Telescope. 
Limiting Doppler precisions of 3\,\ms\ have been demonstrated 
from the first 2.5 years of operation, making this the highest precision
planet search in the southern hemisphere. 
From these data we report results for two new sub-stellar
detections. The first is a ``51\,Peg''-like planet around the star 
HD\,179949 with M\,$\sin i = 0.84$\,\Mjup. Photometric study reveals 
this is not a transiting system.
The second is a brown dwarf or very low-mass star companion to 
HD\,164427 in an  eccentric orbit with M\,$\sin i = 46$\,\Mjup.
Hipparcos data indicate this latter object is unlikely to have a mass 
greater than 0.18\Msol.

\end{abstract}


\section{Introduction}

Since the discovery of the first extra-solar planet by Doppler velocity
techniques in 1994 by \citet{mq95}, planetary detections have been dominated 
by northern hemisphere search programmes -- most prolifically by the precision velocity
programmes at Lick (e.g. \citet{bmwmd96}) and Keck (e.g. \citet{vmba00}), but 
also complemented by lower precision
programmes at OHP \citep{elodie}, McDonald Observatory \citep{md97},
AFOE \citep{afoe97}, and programmes at La Silla \citep{eso,coralie}.
Of these programmes, only the latter have access to the sky south of $\sim -20$\arcdeg, and
these achieve precisions of $\sim$10\,\ms. In 1998, therefore, the Anglo-Australian
Planet Search was begun to complete the all-sky coverage of the brightest stars
at precisions reaching 3\,\ms.
In this paper we present some first results from this programme. A companion paper
\citep[hereafter Paper II]{btmjpa00} presents results for a further two new planets,
along with a detailed description of our observational programme.

\section{The Anglo-Australian Planet Search}

The Anglo-Australian Planet Search is being carried out on the 3.92\,m Anglo-Australian
Telescope (AAT), using the University College of London Echelle Spectrograph (UCLES)
and an I$_2$ absorption cell.
UCLES is operated in its 31\,lines\,mm$^{-1}$ mode with an MIT/LL 2048$\times$4096 15$\mu$m pixel
CCD. This lumogen-coated CCD (denoted MITLL2) underwent a serious failure at the end of
1999, which required its read-out to be switched to the second of its working amplifiers
in 2000 January (following which it was denoted MITLL2a). No change in the operation 
of the detector (apart from a flipped read-out format) has been detected by our programme.

Our target sample of $\approx$200 stars with $\delta < -20$\arcdeg, includes F,G and K V-IV stars with V$<$7.5
and MV stars with V$<$11.5. Where age/activity information is available 
from S or $R_{hk}$ indices (see e.g., \citet{hsdb96}), target stars are required to have ages
greater than 3\,Gyr. Our first observing run was in 1998 January, and the last run for which
observations are reported here was in 2000 November. The observing and data processing 
procedure follows that described in \citet{bmwmd96}, and is described in 
detail in Paper II. In particular, Paper II presents velocities for a number of stable
stars illustrating that we reach a velocity precision floor of 3\,\ms\ for the
bright stars included in our programme.
One of the stars presented here (HD\,179949) has also had confirmatory observations
acquired at three epochs in 2000 September 5-8, as part of the Keck planet 
search \citep{vmba00}.

\section{Characteristics of HD\,179949 \& HD\,164427}

\begin{figure}
\plotone{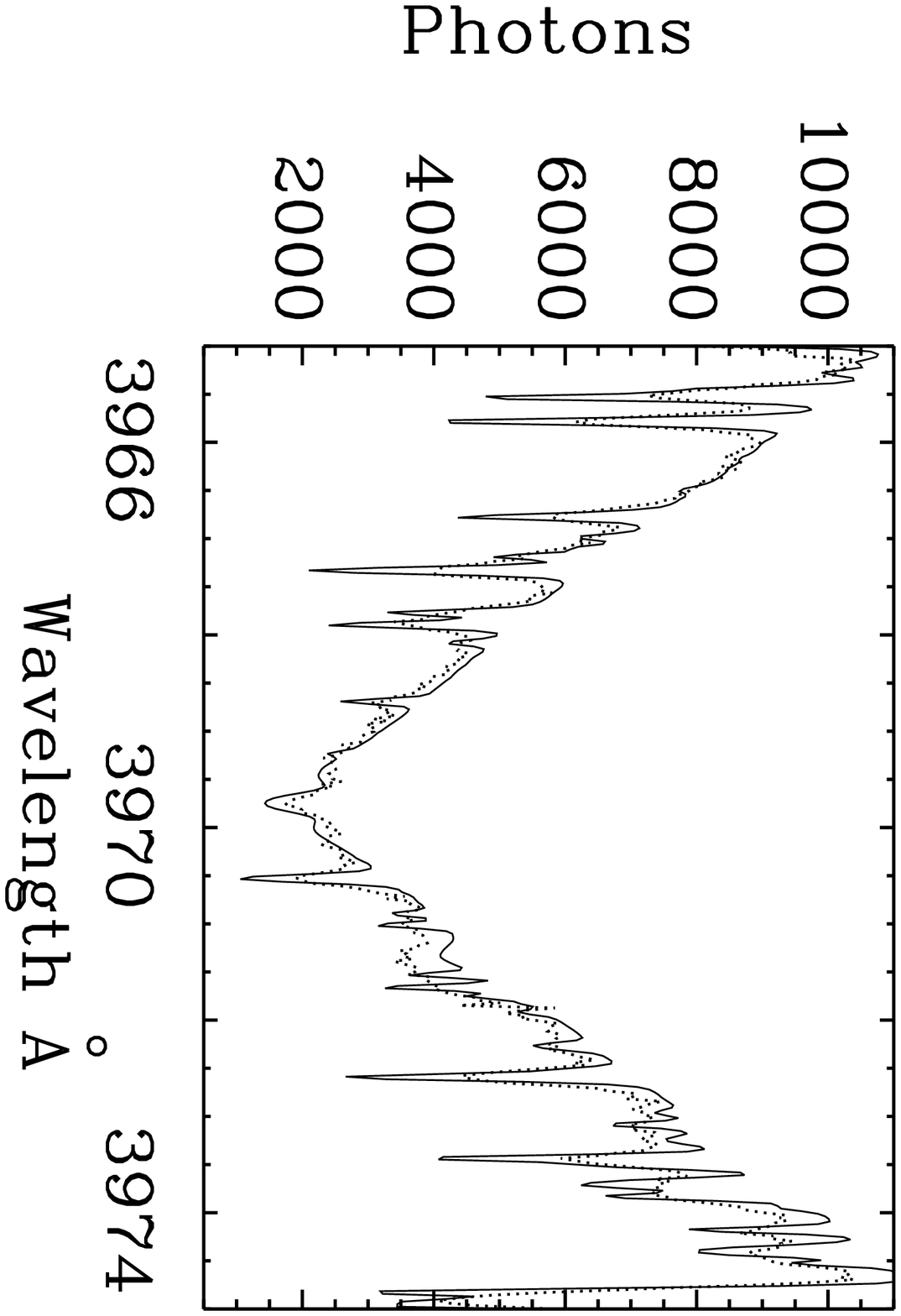}
\caption{Comparison of the \ion{Ca}{2} H line core in the Sun (solid line)
   and HD\,179949 (dotted line).  With $R'_{hk}$ = $-$4.72 
   and $v\,\sin i$=6.3\,\kms, HD\,179949 is
   slightly more active than the Sun and rotates womewhat faster.  The estimated Doppler
   velocity ``jitter'' in an F8V star due to this $\approx$10\,\ms\ (see text).}
\label{hd179949_hline}
\end{figure}

HD\,179949 (HR\,7291, HIP\,94645, GJ749) is an F8V star, which shows moderate rotation
with a measured $v \sin i$ = 6.3$\pm$0.9\,\kms \citep{gpv96}. Its Hipparcos 
parallax puts it at 27.0$\pm$0.5\,pc, with M$_V$=4.09$\pm$0.04 \citep{esa97}.
%
%
There is no published evidence to indicate that HD\,179949 is a binary.
The photometry of \citet{eggen97}
indicates HD\,179949 has roughly solar metallicity with [Fe/H]=+0.02$\pm$0.1.
The $ubvy$ calibration of \citet{vmba00} and published photometry
would indicate a more metal-rich [Fe/H]=+0.22$\pm$0.07.
The latter is supported by \citet[Fig. 8]{tt95} which indicates
HD\,179949 is as metal-rich, if not more so, than the majority of disk main-sequence stars.
HD\,179949 has a moderate X-ray luminosity as determined by ROSAT's PSPC
 -- \citet{hsv98} determine a L$_X$=41.0$\times$10$^{27}$ erg\,s$^{-1}$, while
in an independent analysis \citet{pvs98} determine 
L$_X$/L$_{bol}$=(9.7$\pm$3.3)$\times$10$^{-6}$.
Both are some 10 times higher than the equivalent quantity in the quiet Sun.
The spectra acquired at Keck have been used to derive a  $R'_{hk}$ index
of -4.72 for this star, using the same procedure as \citet{vmba00}.
Figure \ref{hd179949_hline} compares the \ion{Ca}{2} H line with that seen in
the Sun.
%
%
The mass of HD\,179949 is estimated to be 1.24$\pm$0.1\Msol, based on interpolation
between evolutionary tracks of \citet{fpb97,fpb98}.

\placefigure{hd179949_hline}  

%
%

HD\,164427 (HIP\,88531, 
Gl\,700A) is an  inactive G0V \citep{mssI} star, with a $R'_{hk}$ index of {-4.95}
\citep{hsdb96}. Its Hipparcos parallax puts it at a distance of 39.1$\pm$1.4\,pc,
and M$_V$=3.91$\pm$0.8 \citep{esa97} making it somewhat over-luminous for its
spectral type. Indeed \citet{elms64} classified it as a sub-giant, with luminosity class IV.
The $uvby$ calibration of \citet{vmba00} suggests a metallicity of [Fe/H]=+0.11$\pm$0.07.
HD\,164427 was catalogued as a very wide binary (Gl\,700AB) by \citet{gliese69} 
(though both have since dropped out of
the nearby star sample \citep{gj91} due to better parallaxes placing them 
outside the 25\,pc limit).  The binary identification 
is actually due to Luyten (1957) who catalogued this pair as LTT\,7172
and LTT\,7173 (respectively) with common proper motion and
a 28\arcsec\ separation at a position angle of 336\arcdeg. There is no subsequent
astrometry of this pair in the literature. Examination of 
Digital Sky Survey scans of a UKST plate from 1992.6 shows no evidence for 
an object with the magnitude difference indicated by Luyten ($\Delta\,m = 7.2$)
at this separation relative to HD\,164427, leading us to conclude it may
not be a common proper motion pair. Even if the system were a binary, it would be
a very wide system (1090\,au) -- so wide as to be irrelevant
for the purposes of high precision doppler velocities.
%
%
%
%
The mass of HD\,164427 is estimated to be 1.05$\pm$0.1\Msol.

Both stars were seen to be photometrically stable over the life of the Hipparcos mission
at a 95\% confidence level of $<$0.015 magnitudes \citep{esa97}.

\section{Radial Velocity Observations and Orbital Solutions}

\begin{figure}
\plotone{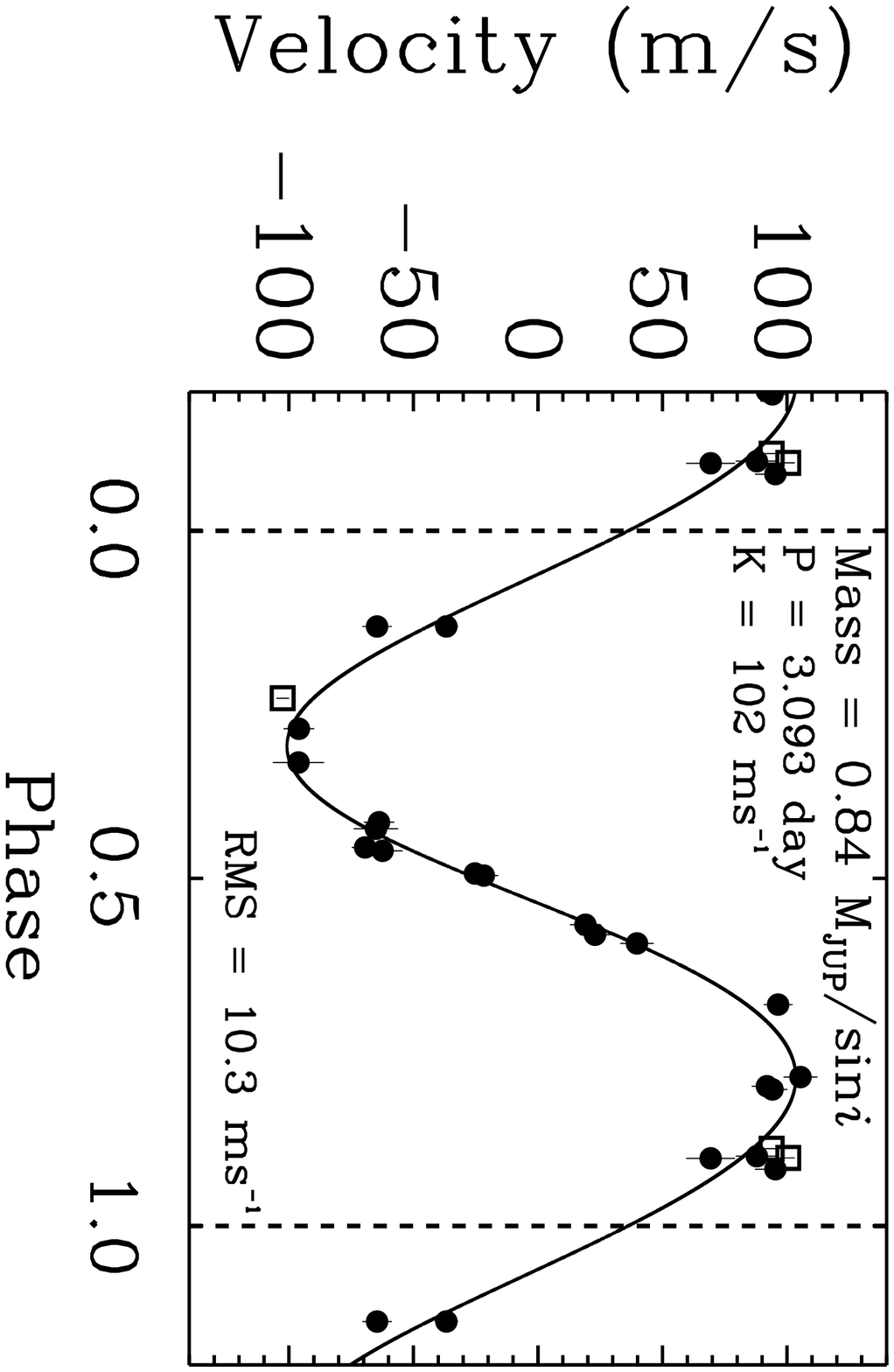}
\caption{AAT {\em (dots)} and Keck {\em (squares)} Doppler velocities for HD\,179949 phased with a period of 
3.093 days.  
The rms of the velocities about the fit is 10.3\,\ms. The solid line is a best
fit Keplerian with the parameters shown
in Table \ref{orbits}.
Assuming 1.24\,\Msol\ for the primary,
the minimum (\Msini) mass of the companion is 0.84$\pm$0.05\,\Mjup, and
the semimajor axis is 0.045$\pm$0.004\,au.}
\label{hd179949_rv_curve}
\end{figure}

Twenty--three observations of HD\,179949 are listed in Table \ref{vel179949}, where
the column labelled ``Uncert.'' is the velocity uncertainty produced by the
least-squares fitting process, which simultaneously determines the Doppler shift and
the spectrograph point-spread function (PSF), given an iodine absorption spectrum,
an ``iodine free'' template spectrum of the object, and an ``iodine'' spectrum 
of the object \citep{bmwmd96}. This uncertainty includes the effects of photon-counting uncertainties,
residual errors in the spectrograph PSF model and variation in the underlying 
spectrum between the template and ``iodine'' epochs. Only observations
where the uncertainty is less than twice the median uncertainty are
listed.
These data are shown in Figure \ref{hd179949_rv_curve} phased with a period of 3.093 days. 
The figure shows the best-fit Keplerian model for the
data, with the resultant orbital parameters listed in Table \ref{orbits}. 
The best--fit period from Keplerian fitting
is the same as the period found from the peak of the Scargle periodogram, with a false
alarm probability less than 0.001. Both AAT (dots) and Keck (squares) data are shown, with the Keck data
completely confirming the orbital fits derived from the AAT data alone.
Like the previously observed giant planets in 3 to 5 day orbits,
HD\,179949 appears to be in a circular orbit.  Table \ref{orbits} includes the best--fit
orbital solutions both for the case of a circular orbit and floating eccentricity.
The resulting minimum companion mass is 0.84\Mjup, with an orbital semi-major axis
of 0.045\,au.

The residuals about the fit are at 
the level of ``jitter'' expected in an F8V star with HD\,179949's level
of rotation (6.3\,\kms) and  activity ($R'_{hk}$=$-4.72$).
``Jitter'' here is used to refer to the scatter in the observed
velocity about a mean vaue in systems observed over the long-term to 
have no Keplerian Doppler shifts. It is thought to be the combined
effect of surface inhomogeneties, stellar activity and stellar rotation.
\citet{sbm98} have studied the correlations between ``jitter'',
and  stellar rotation and activity in the Lick precision velocity programme. 
They indicate we would expect HD\,179949
to show $\approx$10\,\ms\ ``jitter'' due to activity, spots, etc. We measure a scatter 
of 10\,\ms\ about our Keplerian fit, which is consistent with this expectation.

\placetable{vel179949}  

\placefigure{hd179949_rv_curve} 

\placetable{orbits}  

\begin{figure}
\plotone{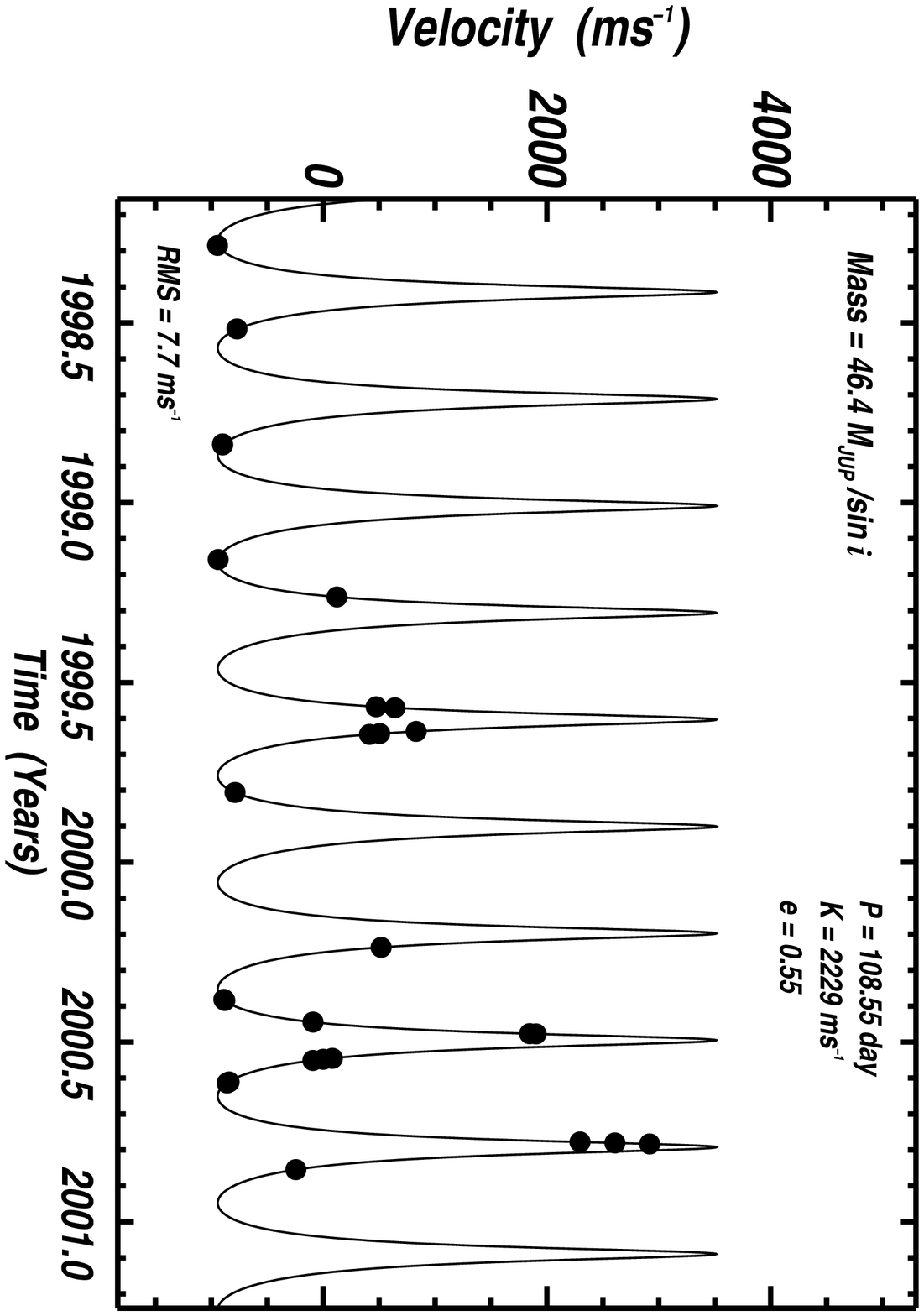}
\caption{AAT Doppler velocities for HD\,179949 from 1998 January to
2000 September. The solid line is a best fit Keplerian with the parameters shown
in Table \ref{orbits}.
The rms of the velocities about the fit is 7.7\,\ms. Assuming a 1.05\,\Msol\ for the primary,
the minimum (\Msini) mass of the companion is 46.4$\pm$3.4\,\Mjup, and
the semimajor axis is 0.46$\pm$0.05\,au.}
\label{hd164427_rv_curve}
\end{figure}

The twenty-seven observations of HD\,166427 are listed in Table \ref{vel164427}, and they are shown in Figure \ref{hd164427_rv_curve}
along with a Keplerian fit to the data with
the orbital parameters listed in Table \ref{orbits}. The rms scatter about this fit 
is 7.7\,\ms, slightly larger than the median uncertainty from the fitting process
of 5.6 \ms. This is
larger than the systematic precision limit of the Anglo-Australian Planet Search,
as the earliest observations of this object were performed in very poor conditions.
As soon as it was realised HD\,164427 was a velocity variable
with large amplitude, it was ``pro-rated'' within the programme to only short 
exposures (and so lower precision than the rest of the programme), though still more 
than sufficient to precisely
determine the companion's orbit.
The resultant minimum companion mass is 46\Mjup, or 0.043\Msol, and the
orbital semi-major axis is 0.46\,au.

\placetable{vel164427}  

\placefigure{hd164427_rv_curve} 

\section{Discussion}

\subsection{HD\,179949}

Ten other ``51 Peg''-like planets are currently known with periods under 10 days,
and eccentricities less than 0.1\footnote{In the absence of a formal classification
scheme, we consider HD\,83443, HD\,168746, HD\,46375, HD\,75289,
BD-10\,3166, HD\,187123, HD\,209458, $\upsilon$\,And\,b, $\tau$\,Boo to be ``51\,Peg''-like, as well as 51\,Peg itself.}. The minimum masses of these planets span the range 0.24-3.87\,\Mjup,
with orbital semi-major axes in the range 0.038-0.059\,au.
In all respects, HD\,179949 falls squarely into this class of objects. Even its
above-solar metallicity is similar to that seen in the other ``51\,Peg''-like planets 
(e.g., \citet{coralie}, \citet{gonz00}).

\subsection{Transit Search in HD\,179949}

As with all the other ``51\,Peg''-like planets, HD\,179949 is an excellent
candidate for a transit search.  The Hipparcos magnitude, color index, and
parallax of the star imply a radius of about 1.24 R$_{\odot}$, which,
combined with the orbital radius of 0.045 au, places the probability of
observable transits at 12.8\%.  Successful transit observations (already 
carried out for HD\,209458 by \citet{cblm00} and \citet{hmbv00}) not only 
tightly constrain orbital inclination, but provide planetary radii as well.

To search for transits in HD\,179949, we obtained photometric observations 
between 2000 September 21 and October 1 with the T8 0.80\,m Automatic
Photoelectric Telescope (APT) at Fairborn Observatory in southern 
Arizona\footnote{These observations are available at 
{\tt http://schwab.tsuniv.edu/t8/hd179949/hd179949.html}.}.  The instrumentation and
techniques used for this transit search are described in \citet{h99}
and \citet{hbdfs00}.  The observations were
made differentially with respect to the comparison star HD\,178075 ($V = 6.30$,
$\bv = 0.01$, B9.5V).  The precision of an individual observation is
approximately 0.004\,mag, somewhat worse than the typical precision of
0.001\,mag obtainable with the APT due to observations being at air masses between two
and three.

No transit events were detected at the times predicted by the orbital
parameters of Table \ref{orbits}.  For an assumed planetary radius of 1.4\,\Rjup\
(the measured radius of the planetary companion of HD\,209458 \citep{hmbv00,j00}),
the predicted transit depth is 0.014\,mag.  The mean of our 16 photometric 
observations of HD\,179949 taken within the predicted transit window is
0.1058 $\pm$ 0.0010\,mag; the mean of the 38 observations outside the
transit window is 0.1044 $\pm$ 0.0007\,mag.  Thus, the two means agree within
their respective errors and transits deeper than about 0.001\,mag are ruled 
out.  This non-detection of transits limits the orbital inclination $i$
to less than 83\arcdeg, and $\sin i$ to less than 0.992.  Furthermore,
the photometric observations place a limit of about 0.001\,mag on any
photometric variability of HD\,179949 on the radial velocity period.  This 
confirms that neither starspots nor stellar pulsations can be the cause of 
the radial velocity variations (see \citet{hbdfs00}) and so
strongly supports the existence of the planet even in the absense of 
transits.

\subsection{HD\,164427}

Recent work on the mass function of brown dwarfs in star 
   clusters \citep{letal00, lr00, z00} has failed 
   to show that star formation processes cannot form objects below the
   13\,\Mjup\ deuterium burning limit. (Though the same results can
   be interpreted - less straightforwardly - as implying the existence
   of free-floating sub-13\,\Mjup\ objects formed by dynamical evolution
   ejecting objects formed by planetary processes.) Certainly
   they provide no evidence that the deuterium burning limit is a useful
   demarcation boundary between star- and planet-formation processes.
   Difficult as it may be, only an understanding of formation mechanisms 
   (and possible subsequent dynamical evolution) can allow us to determine 
   whether objects in the 5-20\,\Mjup\ range
   are planets or brown dwarfs.
Nonetheless, the minimum mass of the companion to HD\,164427 
clearly lies in a range unlikely to be due to planetary formation processes -- it 
is a brown dwarf or a star, not a planet.

Hipparcos astrometry, however, allows us to place some limits on the orbital inclination
and mass for HD\,164427's companion. The astrometric solution for HD\,164427 shows a good
fit for the parallax and proper motion of a single star, with 1-$\sigma$ residuals of
$\approx$3\,mas. The uncertainty on the derived parallax is 0.9\,mas. Based on this
we can assume an upper limit to the astrometric perturbation ($a_0$) of HD\,164427 due to an
unseen companion of $\approx 2$\,mas. Then from the relationship between the 
spectroscopic and astrometric orbital elements (e.g. \citet{hamuq00}, equation 1)
we 
conclude that $\sin i > 0.24$, or equivalently that the companion mass is less 
than 190\,\Mjup\ or 0.18\,\Msol, making it either a brown dwarf, or a very low-mass star. 
With an orbital semi-major axis of 0.46\,au at 39\,pc, such a companion will
have an maximum apparent separation of 11.8\,mas. Table \ref{companion_mag} compares
the absolute K magnitudes of HD\,164427, with predicted magnitudes of the companion
in the range of allowed masses. Only at the very massive end of this range is the
companion likely to be detectable by either ground-based adaptive optics or space-based
imaging.

Nonetheless, such a challenging observation must be pursued. Of the small number
of brown dwarf candidates  identified by precision Doppler surveys to date (around 11), many
have been shown to actually be stars \citep{hamuq00}, making {\em bona fide}
brown dwarf companions rare, and well worth unambiguously identifying.

\section{Conclusions}

We present first results from the Anglo-Australian Planet Search, which is
now the highest precision Doppler planet search in the southern hemisphere. These 
include a new member of the ``51\,Peg''-like class of planets around the
F8 dwarf HD\,179949, and 
a new brown dwarf candidate companion to the G0 dwarf HD\,164427. This programme has now
been running on the AAT for over 2 years and has demonstrated velocity precisions reaching
3\,\ms. We confidently expect many more discoveries in the years to come -- particularly as it 
extends its sensitivity toward the Jupiter-like planet regime.

\acknowledgments
The Anglo-Australian Planet Search team would like to gratefully acknowledge the support
of the Director of the AAO, Dr Brian Boyle, and the superb technical support which has been
received throughout the programme from AAT staff - in particular E.Penny, R.Patterson, D.Stafford, F.Freeman, S.Lee, J.Pogson and G.Schafer. We further acknowledge support 
by; the partners of the Anglo-Australian Telescope Agreement (CGT,HRAJ,AJP);
    NASA grant NAG5-8299 \& NSF grant AST95-20443 (GWM);
    NASA grant NAG5-4445 \& NSF grant AST-9619418 (SSV);
    NSF grant AST-9988087 (RPB); NASA grants NCC5-96 \& NCC5-228, 
    NSF grant HRD-9706268  \& the Richard Lounsbery Foundation (GWH); and
    Sun Microsystems.

\clearpage

\clearpage

\begin{deluxetable}{lrr}
\tablenum{1}
\tablecaption{Velocities for HD\,179949}
\tablewidth{0pt}
\tablehead{
JD & RV & Uncert \\
(-2440000)   &  (\ms) & (\ms)
}
\startdata
  11120.910   & 118.1  &  5.8 \\
  11383.024   & -43.3  &  8.9 \\
  11410.943   & -47.7  &  5.3 \\
  11413.052   & -15.0  &  4.7 \\
  11413.924   & -42.1  &  6.1 \\
  11472.917   &  -3.6  &  5.7 \\
  11683.133   & -40.6  &  8.0 \\
  11684.180   & 113.7  &  6.1 \\
  11706.151   &  91.1  &  9.8 \\
  11718.161   & 127.1  &  6.8 \\
  11742.960   & 115.9  &  5.9 \\
  11743.992   & -42.9  &  5.9 \\
  11745.101   &   0.0  &  5.8 \\
  11766.970   &  40.8  &  6.2 \\
  11768.000   & 109.6  &  8.5 \\
  11770.107   &  44.6  &  6.1 \\
  11792.751\tablenotemark{a}  & 122.2  &  2.6 \\
  11793.798\tablenotemark{a}  & -80.7  &  2.5 \\
  11795.803\tablenotemark{a}  & 115.6  &  2.5 \\
  11827.956   & -74.3  &  6.1 \\
  11828.911   &  61.5  &  6.7 \\
  11829.915   & 117.1  &  8.2 \\
  11855.942   & -74.4  & 10.3 \\
\enddata
\tablenotetext{a}{Observations from Keck. All other observations are from the AAT.}
\label{vel179949}
\end{deluxetable}

\clearpage

\begin{deluxetable}{lccc}
\tablenum{2}
\tablecaption{Orbital Parameters}
\tablewidth{0pt}
\tablehead{
\colhead{Parameter}            & \colhead{HD\,179949} 
                                                   & \colhead{HD\,179949}
                                                                      &\colhead{HD\,164427} \\
                               & ($e$=0.0)         & ($e$ fitted)     &
}
\startdata
Orbital period $P$ (d)          & 3.093$\pm$0.001  & 3.093$\pm$0.001   & 108.55$\pm$0.04\\
Velocity amp. $K$ (\ms)         &  101.3$\pm$3.0    & 102.2$\pm$3.0     &   2229$\pm$77.0  \\
Eccentricity $e$                &    0.0 (fixed)    & 0.05$\pm$0.03     & 0.55$\pm$0.02  \\
$\omega$ (\arcdeg)              &    0.0            & 226$\pm$25        &356.9$\pm$0.5   \\
$a_1 \sin i$ (km)               &  4307.3$\pm$15.0  & 4338.7$\pm$15.0  &(2.776$\pm$0.049)$\times$10$^6$\\
Periastron Time (JD-244000)     & 11718.19$\pm$0.1 &11723.24$\pm$0.1  &11724.6$\pm$0.2 \\
\Msini\ (\Mjup)                   &  0.84$\pm$0.05    &  0.84$\pm$0.05    & 46.4$\pm$3.4\\
a (AU)                          &  0.045$\pm$0.004  &  0.045$\pm$0.004  & 0.46$\pm$0.05\\
RMS about fit (\ms)             &  10.8             &  10.3             & 7.7    \\
\enddata
\label{orbits}
\end{deluxetable}

\clearpage

\begin{deluxetable}{lrr}
\tablenum{3}
\tablecaption{Velocities for HD\,164427}
\tablewidth{0pt}
\tablehead{
JD & RV & Uncert \\
(-2440000)   &  (\ms) & (\ms)
}
\startdata
 10917.287  & -1071.2  &  7.5 \\
 11002.091  &  -897.7  & 10.2 \\
 11118.891  & -1026.5  &  9.8 \\
 11119.905  & -1025.7  &  8.3 \\
 11236.281  & -1065.7  &  6.4 \\
 11274.300  &    -3.9  &  5.3 \\
 11385.868  &   347.2  &  6.8 \\
 11386.879  &   513.8  &  4.9 \\
 11410.911  &   703.9  &  4.2 \\
 11413.034  &   378.0  &  3.6 \\
 11413.904  &   286.7  &  5.6 \\
 11472.893  &  -914.8  &  4.6 \\
 11630.300  &   391.3  &  4.9 \\
 11683.096  & -1014.7  &  5.6 \\
 11684.147  & -1006.3  &  6.4 \\
 11706.112  &  -217.9  &  4.9 \\
 11717.923  &  1718.7  &  5.3 \\
 11718.123  &  1775.8  &  4.9 \\
 11742.913  &   -44.6  &  5.5 \\
 11743.933  &  -126.6  &  7.0 \\
 11745.067  &  -218.7  &  4.6 \\
 11766.942  &  -968.3  &  5.2 \\
 11767.992  &  -985.4  &  8.4 \\
 11827.914  &  2169.0  &  5.5 \\
 11828.890  &  2480.8  &  6.3 \\
 11829.894  &  2791.5  &  7.0 \\
 11855.921  &  -371.0  &  8.0 \\
\enddata
\label{vel164427}
\end{deluxetable}

\clearpage

\begin{deluxetable}{llcc}
\tablenum{4}
\tablecaption{Predicted K magnitudes for HD\,164427 companion}
\tablewidth{0pt}
\tablehead{
Companion Mass & Age &M$_{K}$ & $\Delta$M$_{K}$}
\startdata
\multicolumn{3}{c}{HD\,164427 M$_K$=5.4} \\
0.18\,\Msol\tablenotemark{a}& \nodata& 7.7  & 2.3\\
0.10\,\Msol\tablenotemark{a}& \nodata& 9.2  & 3.8\\
75\,Mjup\tablenotemark{b}   & 1\,Gyr & 11.2 & 5.9\\
            & 5\,Gyr & 11.7 & 6.3\\
42\,Mjup\tablenotemark{b}   & 1\,Gyr & 13.7 & 8.3\\
            & 5\,Gyr & $>$16&$>$10.6 \\
\enddata
\tablenotetext{a}{Empirical M$_K$ values due to \citet{hm93}.}
\tablenotetext{b}{Model M$_K$ values due to \citet{cbah00}.}
\label{companion_mag}
\end{deluxetable}

\end{document}